\documentclass{nature}

\usepackage{amsmath}
\usepackage{array}
\usepackage{gensymb}
\usepackage{url}
\usepackage{xspace}
\usepackage[utf8]{inputenc}
\usepackage{microtype}
\usepackage{ulem}
\usepackage{multibib}
    \newcites{mt,me}{References for main text, References for method}

\usepackage[dvipsnames]{xcolor}

\usepackage{graphicx}
\makeatletter
\let\saved@includegraphics\includegraphics
\AtBeginDocument{\let\includegraphics\saved@includegraphics}
\renewenvironment*{figure}{\@float{figure}}{\end@float}
\makeatother

\newcolumntype{L}{>{$}l<{$}} 

\newcommand{\phx}{Phoenix}
\newcommand{\SSSSS}{$S^5$\xspace}
\newcommand{\gaia}{\textit{Gaia}\xspace}
\newcommand{\feh}{\ensuremath{[\textrm{Fe}/\textrm{H}]}}
\newcommand{\kms}{\ensuremath{\textrm{km}\,\textrm{s}^{-1}}}

\title{\Huge The tidal remnant of an unusually\\ metal-poor globular cluster}

\author{Zhen~Wan$^{1}$,
Geraint~F.~Lewis$^{1}$,
Ting~S.~Li$^{2,3,4,5,6}$,
Jeffrey~D.~Simpson$^{7}$,
Sarah~L.~Martell$^{7,8}$,
Daniel~B.~Zucker$^{9,10}$,
Jeremy~R.~Mould$^{11}$,
Denis~Erkal$^{12}$,
Andrew~B.~Pace$^{13}$,
Dougal~Mackey$^{14}$,
Alexander~P.~Ji$^{2,15}$,
Sergey~E.~Koposov$^{13,16}$,
Kyler~Kuehn$^{17,18}$,
Nora~Shipp$^{19,5,4}$,
Eduardo~Balbinot$^{20}$,
Joss~Bland-Hawthorn$^{1,8}$,
Andrew~R.~Casey$^{21}$,
Gary~S.~Da~Costa$^{14}$,
Prajwal~Kafle$^{22}$,
Sanjib~Sharma$^{1,8}$
\&
Gayandhi~M.~De~Silva$^{18,8}$
(\SSSSS Collaboration)
}

\begin{document}
\maketitle
{\scriptsize
\begin{affiliations}
\item Sydney Institute for Astronomy, School of Physics, A28, The University of Sydney, NSW 2006, Australia
\item Observatories of the Carnegie Institution for Science, 813 Santa Barbara St., Pasadena, CA 91101, USA
\item Department of Astrophysical Sciences, Princeton University, Princeton, NJ 08544, USA
\item Fermi National Accelerator Laboratory, P.O.\ Box 500, Batavia, IL 60510, USA
\item Kavli Institute for Cosmological Physics, University of Chicago, Chicago, IL 60637, USA
\item NHFP Einstein Fellow
\item School of Physics, UNSW, Sydney, NSW 2052, Australia
\item Centre of Excellence for All-Sky Astrophysics in Three Dimensions (ASTRO 3D), Australia
\item Department of Physics \& Astronomy, Macquarie University, Sydney, NSW 2109, Australia
\item Macquarie University Research Centre for Astronomy, Astrophysics \& Astrophotonics, Sydney, NSW 2109, Australia
\item Centre for Astrophysics and Supercomputing, Swinburne University of Technology		PO box 218 Hawthorn Vic 3122 Australia
\item Department of Physics, University of Surrey, Guildford GU2 7XH, UK
\item McWilliams Center for Cosmology, Carnegie Mellon University, 5000 Forbes Ave, Pittsburgh, PA 15213, USA
\item Research School of Astronomy and Astrophysics, Australian National University, Canberra, ACT 2611, Australia
\item Hubble Fellow
\item Institute of Astronomy, University of Cambridge, Madingley Road, Cambridge CB3 0HA, UK
\item Lowell Observatory, 1400 W Mars Hill Rd, Flagstaff,  AZ 86001, USA
\item Australian Astronomical Optics, Faculty of Science and Engineering, Macquarie University, Macquarie Park, NSW 2113, Australia
\item Department of Astronomy \& Astrophysics, University of Chicago, 5640 S Ellis Avenue, Chicago, IL 60637, USA
\item Kapteyn Astronomical Institute, University of Groningen, Landleven 12, 9747 AD Groningen, The Netherlands
\item School of Physics and Astronomy, Monash University, Wellington Rd, Clayton 3800, Victoria, Australia 
\item ICRAR, The University of Western Australia, 35 Stirling Highway, Crawley, WA 6009, Australia
\end{affiliations}
}

\newpage 

\begin{abstract}
Globular clusters are some of the oldest bound stellar structures observed in the Universe\citemt{1991ARA&A..29..543H}. They are ubiquitous in large galaxies and are believed to trace intense star formation events and the hierarchical build-up of structure\citemt{2006ARA&A..44..193B,2019Natur.574...69M}. Observations of globular clusters in the Milky Way, and a wide variety of other galaxies, have found evidence for a `metallicity floor', whereby no globular clusters are found with chemical (`metal') abundances below approximately 0.3 to 0.4 per cent of that of the Sun\citemt{1996AJ....112.1487H,2018RSPSA.47470616F,2019MNRAS.487.1986B}. The existence of this metallicity floor may reflect a minimum mass and a maximum redshift for \textit{surviving} globular clusters to form, both critical components for understanding the build-up of mass in the universe\citemt{2019MNRAS.486L..20K}.
Here we report measurements from the Southern Stellar Streams Spectroscopic Survey of the spatially thin, dynamically cold \phx\ stellar stream in the halo of the Milky Way. The properties of the \phx\ stream are consistent with it being the tidally disrupted remains of a globular cluster. However, its metal abundance ($\feh=-2.7$) is substantially below that of the empirical metallicity floor.
The \phx\ stream thus represents the debris of the most metal-poor globular cluster discovered so far, and its progenitor is distinct from the present-day globular cluster population in the local Universe. Its existence implies that globular clusters below the metallicity floor have probably existed, but were destroyed during Galactic evolution.
\end{abstract}

The \phx\ stellar stream is a thin over-density of stars in the Milky Way halo. It spans approximately 8$^{\circ}$ lengthwise on the sky and was originally identified in the Dark Energy Survey (DES)\citemt{2016ApJ...820...58B}. 
Comparison of the DES photometry with theoretical isochrones shows that the stream to be located at a heliocentric distance of about 19 kpc, and that its constituent stars are old and metal-poor, although the isochrone fits do not allow precise determination of these quantities\citemt{2018ApJ...862..114S}.
The narrow width of the stream in the plane of the sky (about 50 pc) suggests that the progenitor was a low-mass Milky Way satellite (mass $M\sim 3\times 10^4 M_\odot$, where the $M_\odot$ is the mass of the Sun)\citemt{Erkal:2016,2018ApJ...862..114S}, which has now been completely disrupted. Dynamical modelling has revealed that \phx\ is likely part of a much more extensive debris structure that also includes the Hermus stellar stream located around 180$^{\circ}$ away on the sky\citemt{2016ApJ...820L..27G,2016ApJ...830..135C}.

We observed the \phx\ stream as part of the Southern Stellar Streams Spectroscopic Survey (\SSSSS) programme\citemt{Li2019a} to acquire kinematic and chemical abundance measurements along its length. Candidate \phx\ stars were selected by applying broad cuts in both colour-magnitude space (using DES DR1\citemt{Abbott2018} photometry) and proper motion space (using \gaia\ DR2\citemt{2016A&A...595A...1G,2018A&A...616A...1G}). They were observed across seven fields with the 2dF+AAOmega fibre-fed spectrograph on the Anglo-Australian Telescope (AAT). Stellar metallicities and radial velocities were determined by fitting synthetic stellar templates in the region of the Ca{\sc ii} triplet at around 8,600 \hbox{\AA}. Full details of the candidate selection, observations and data reduction are provided in the Methods.

\begin{figure}
    \centering
    \includegraphics[width=0.7\linewidth]{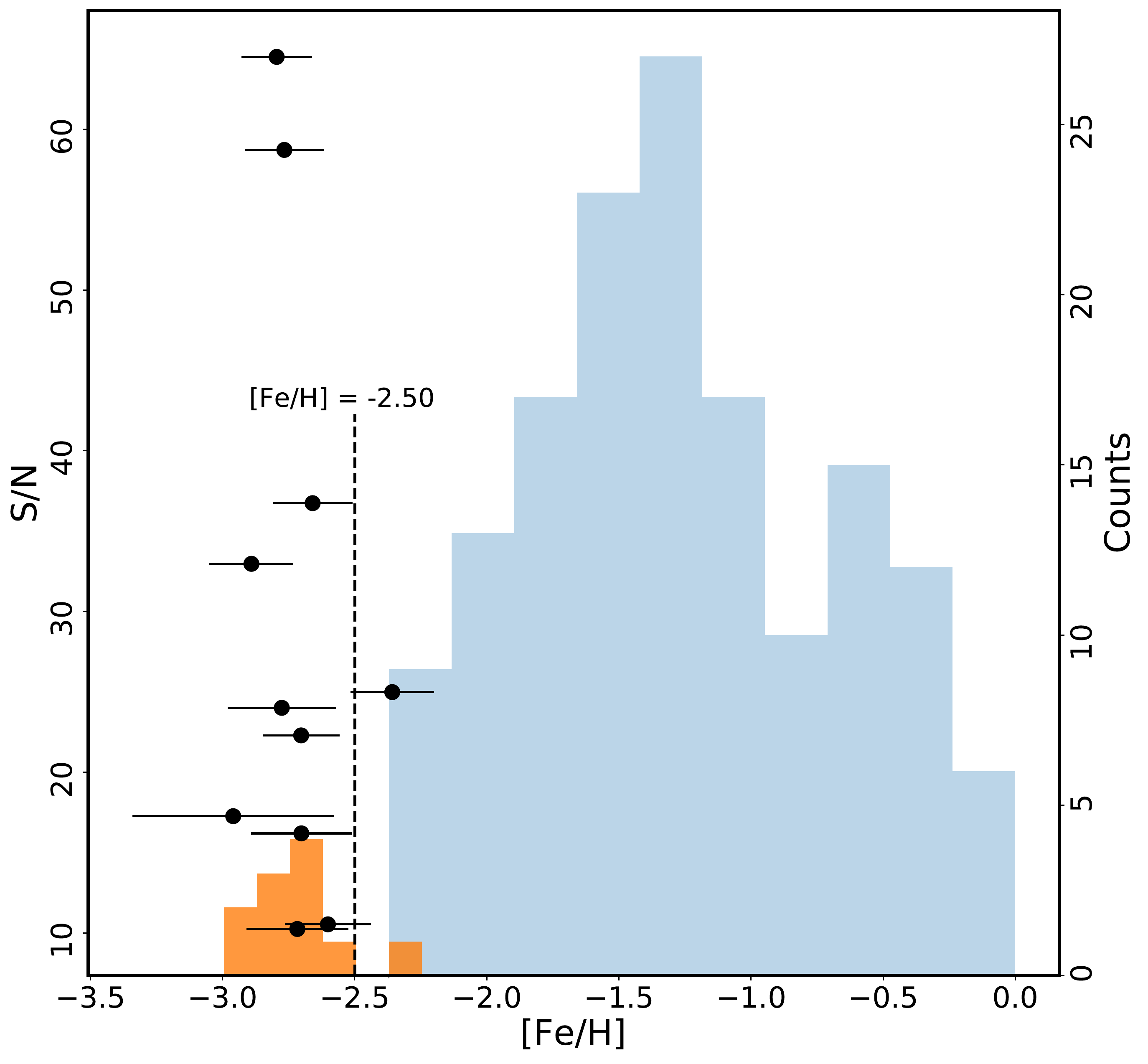}
    \caption{Metallicity versus spectroscopic signal-to-noise ratio for \phx\ stream members. 
    A histogram of the metallicities of \phx\ member stars with signal-to-noise ratios greater than 10 is presented in orange (right axis); the metallicity distribution for individual globular clusters in the Milky Way is shown in blue  \protect\citemt{1996AJ....112.1487H} (right axis). The signal-to-noise ratios of individual \phx\ members are also shown in black points (left axis) with errorbars (1$\sigma$; see ref.\protect\citemt{Li2019a} for detail discussion). The dashed line indicates the location of the empirical `metallicity floor' at $\feh = -2.5$, above which sit all globular clusters in the Milky Way, the Local Group and other nearby galaxies.}
    \label{fig:distmetal}
\end{figure}

In Fig.~\ref{fig:distmetal}(orange histogram) we present the distribution of metallicities for red giant stars in the \phx\ stream that have spectra with a signal-to-noise of more than 10. With one exception the measured metallicities are substantially below $\feh = -2.5$. This is more metal-poor than any known globular cluster in the Milky Way; the metallicity distribution of the Galactic globular cluster population\citemt{1996AJ....112.1487H} is shown with a blue histogram in Fig.~\ref{fig:distmetal}.

\begin{figure}
    \centering
    \includegraphics[width=0.8\linewidth]{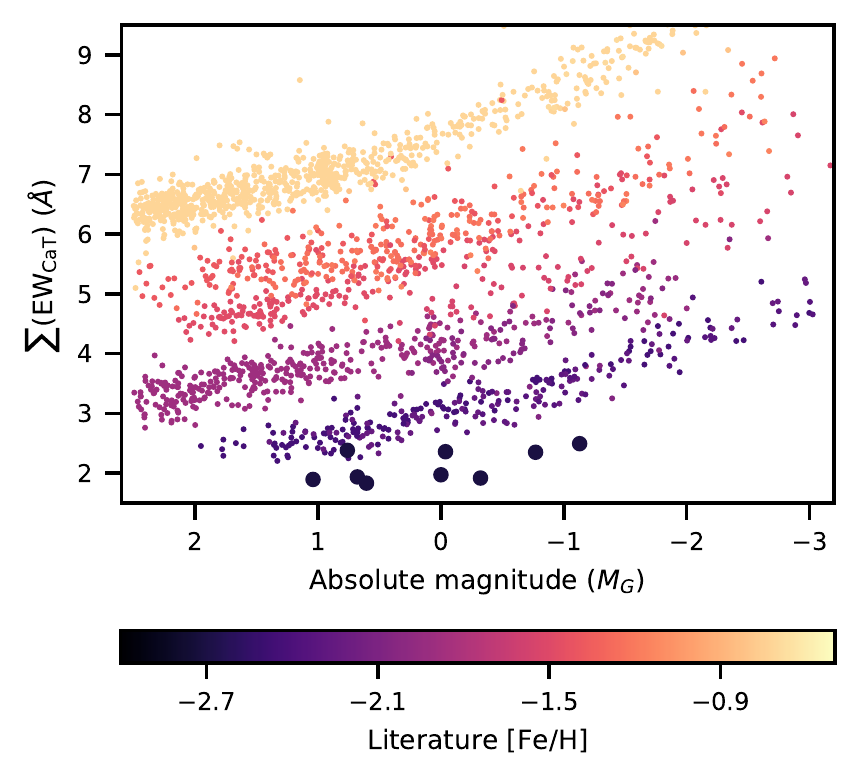}
    \caption{Comparison between the summed equivalent widths of the Ca{\sc ii} triplet. Phoenix members are shown by large circles; the small circles correspond to the 2,050 red-giant-branch stars in 18 other globular clusters observed by the AAOmega spectrograph\protect\citemt{SimpsonJ2020}. For giant branch stars of a given luminosity (that is, absolute magnitude), a smaller equivalent width of the Ca{\sc ii} triplet means that the star has a lower metallicity. The stars are colour-coded by the literature metallicity of their parent globular cluster  \protect\citemt{2019MNRAS.482.1275U}. The lowest-metallicity clusters (those with the weakest Ca{\sc ii} triplet lines for a given absolute magnitude) include M15 (NGC7099; $\feh=-2.44$), M30 (NGC7078; $-2.34$), NGC5053 ($-2.27$), and M68 (NGC4590; $-2.23$). It is clear that the Phoenix members have weaker Ca{\sc ii} triplet lines, and therefore lower metallicities, than all of these clusters.}
    \label{fig:EW}
\end{figure}

To illustrate that this offset is not due to systematic differences between our measurements and those used for the overall compilation, in Fig.~\ref{fig:EW} we present a direct comparison between the summed equivalent widths of the Ca{\sc ii} spectral lines for our \phx\ targets and for 2,050 red giants in 18 Galactic globular clusters spanning a broad metallicity range. Critically, the cluster reference stars were observed using the same facility and instrumental set-up as our \phx\ sample. It is evident that, at a given stellar luminosity, a decreasing equivalent width corresponds to a lower metallicity. The \phx\ members (black circle in Fig.~\ref{fig:EW}) have equivalent widths that are substantially smaller than the most metal-poor cluster in the reference sample, NGC 7099 with $\feh \approx -2.4$, which is among the most metal-poor globular clusters observed in the Milky Way\citemt{2019MNRAS.482.1275U}. 

Figure~\ref{fig:distmetal} suggests that the metallicity spread among our \phx\ sample is comparable to the measurement uncertainties. To quantify this, we used a Markov Chain Monte Carlo (MCMC) approach to explore the joint likelihood space for
mean metallicity and intrinsic dispersion, for the 11 \phx\ stars with a signal-to-noise $\mathrm{S/N} > 10$, given their individual abundance measurements and uncertainties. 
Representing the intrinsic metallicity as a Gaussian, our
analysis yields a mean $\feh = -2.70\pm{0.06}$, and a most-likely intrinsic metallicity dispersion of zero: $\sigma_{\feh} < 0.2$ at 95\% confidence (or $\sigma_{\feh} = 0.07^{+0.07}_{-0.05}$, see Extended Data Fig.~\ref{fig:feh_corner}). This strongly suggests that the \phx\ progenitor comprised a simple stellar population with no self-enrichment in heavy elements.

To further explore the nature of the \phx\ progenitor, we combine our kinematic measurements with dynamical models. After subtracting a polynomial fit for the gradient of the line-of-sight velocity along the stream, we infer a low intrinsic velocity dispersion of $\sigma_v = 2.66_{-0.57}^{+0.72}$\ \kms\ (see Extended Data Fig.~\ref{fig:vgsr_corner}). This is consistent with the idea that the progenitor was a low-luminosity satellite of the Milky Way. To determine its most likely orbit, we integrate numerical models over $3$ billion years in a Milky Way potential including the effect of the Large Magellanic Cloud (LMC), and attempt to reproduce the observed positions on the sky, radial velocities, and proper motions for all the high-likelihood \phx\ members in our sample. As shown in Extended Data Fig.~\ref{fig:sim}, our best-fit model can reproduce these key data, and indicates a prograde orbit with an inclination of about 60$^{\circ}$ relative to the Milky Way disk, a pericenter of approximately 13 kpc, an apocenter of approximately 18 kpc, and an eccentricity of approximately 0.2. The continuation of our stream model passes through the location of the Hermus stream, reinforcing previous suggestions that these two structures are connected\citemt{2016ApJ...820L..27G,2016ApJ...830..135C}.

The narrow on-sky width and small velocity dispersion of the \phx\ stream could only have been produced by a low luminosity globular cluster or an ultra-faint dwarf galaxy (that is, a dwarf galaxy with total luminosity less than $10^5$ times that of the Sun\citemt{Simon2019ARA&A..57..375S}). No other type of system possessing the requisite small size and stellar mass is known. A key distinguishing property for these two classes of object is the internal metallicity spread, which is zero for all except the very brightest globular clusters, but typically larger than about $0.2-0.3$ dex for dwarf galaxies \citemt{Willman2012AJ....144...76W,Simon2019ARA&A..57..375S}; figure 1 of ref.\citemt{Willman2012AJ....144...76W} shows the metallicity spread is 0.3-0.7 dex for 16 dwarf galaxies. Our observation that $\sigma_{\feh} \approx 0$ for \phx\ strongly indicates that the progenitor was a globular cluster. Additional support for this assertion comes from the inferred orbital properties of the stream. It has recently been shown that the ultra-faint dwarfs within $100$\ kpc of the Milky Way have highly eccentric (median $0.8$) and almost exclusively retrograde orbits; the median pericenter distance is around 40 kpc\citemt{Simon2018ApJ...863...89S}.  This is in stark contrast to our preferred trajectory for \phx\, which is much more typical of the orbits inferred for many Galactic globular clusters\citemt{Helmi2018A&A...616A..12G,vasiliev_2019} (see also Extended Data Fig.\ref{fig:actions}).

Consequently, we conclude that the \phx\ stream comprises the tidally disrupted remains of a globular cluster. Our measured mean metallicity, $\feh = -2.70\pm{0.06}$, is thus very notable. Within the Milky Way, no globular cluster is observed to have a metallicity below $\feh \approx -2.5$ (refs.~\citemt{1996AJ....112.1487H,Simpson:2018cm,Simpson2019a}). This empirical metallicity floor extends not only to all other Local Group galaxies\citemt{Larsen:2012cb,2019MNRAS.487.1986B}, but even further, spanning roughly 6 dex in galaxy stellar mass and a wide variety of morphologies and assembly histories\citemt{2018RSPSA.47470616F,2019MNRAS.487.1986B,2019MNRAS.486L..20K}. The \phx\ progenitor therefore apparently occupies a special position, which is distinctly different from the present-day globular cluster population observed in the local Universe.

Theoretical models\citemt{Kruijssen:2015fe,2019MNRAS.486L..20K} point to the galactic mass-metallicity relation at high redshift as the source of the metallicity floor for globular clusters. Galaxies grow through the accretion of gas and other galaxies, and they undergo self-enrichment, creating a correlation between mass and metallicity. At redshift greater than 2, galaxies forming stars with $\feh \approx -2.5$ are predicted to have total stellar masses roughly $10^5-10^6 \mathrm{M_\odot}$. Lower-mass (and hence lower-metallicity) galaxies are unable to form clusters capable of surviving for a Hubble time, resulting in the observed metallicity floor.

It is now well established that the Galactic halo has an underlying smooth, metal poor component (see, for example, ref.\citemt{2019MNRAS.482.3868I}), embellished with non-equilibrium components resulting from more recent infall events. A clear and compelling goal is to identify and describe the most likely collection of building blocks for the Milky Way halo, which has been built up over the age of the Galaxy through accretion and continues to evolve today through the same process.

We have established a low metallicity for the diffuse Phoenix stream, which will continue to dissolve and be absorbed into the ancient smooth halo. It remains unclear whether its metallicity is so low because its progenitor formed in a host galaxy with very low stellar mass ($< 10^5 \mathrm{M_\odot}$), or whether its original host galaxy had a relatively high mass for its metallicity, shielding the \phx\ strean from tidal disruption by the Milky Way until fairly recently. The picture is complicated because the mass of the host galaxy can grow substantially through accretion after the birth of the globular cluster and before it is accreted into the Milky Way.

This result presents two exciting possibilities: first, that additional remnants of metal-poor globular clusters ($\feh < -2.5$) inhabiting the Galactic halo may come to light in future large-scale surveys; and second, that we might associate streams and star clusters brought into the halo in the same accretion event using present-day kinematics and stellar properties. Given our result that the globular cluster with the lowest known metallicity is in the form of a stellar stream rather than an intact self-gravitating system, it is clear that some fraction of the Milky Way halo stars with $\feh < -2.5$ formed in globular clusters.

In support of this hypothesis, we note that two stars in the recently discovered Sylgr stream were found to have $\feh=-2.92\pm0.06$ (ref.~\citemt{Roederer2019}). The nature of the Sylgr progenitor is still unclear. However, combining these observations with our present discovery, and with future surveys targeting low surface-brightness substructures in the Milky Way halo, may yield a new and fuller understanding of the earliest stages of galaxy formation. A test of this scenario is at hand: the James Webb Space Telescope may reveal the association of globular clusters forming along with proto-galaxies in the high redshift universe\citemt{2017MNRAS.469L..63R}.

We note that \phx\ is relatively close to the Palomar 5 stream, the Helmi stream, and the metal-poor globular cluster NGC 5053 in the orbital energy - azimuthal action space, and that \phx\ is spatially well aligned with the Hermus stream. These similarities are discussed further in Methods; although they are intriguing, they are not sufficient to claim a common origin for these systems. Identifying the major components of Galactic halo assembly requires a holistic approach that brings together a number of observations with detailed modelling and probabilistic analysis. Associating a set of globular clusters and streams to a single progenitor galaxy accreted at a particular redshift requires coherence across a range of properties: present-day kinematics and the age-metallicity relation of globular clusters and streams must behave consistently with field stars accreted from the progenitor galaxy. A simultaneous solution for the major progenitors of the Milky Way halo will include progenitor galaxies that are compatible, assembling into a single evolving structure that reproduces the observed halo.

\bibliographystylemt{naturemag}
\bibliographymt{refs}

\newpage
\begin{methods}
The Southern Stellar Stream Spectroscopic Survey (\SSSSS) collaboration was established to measure the kinematics and chemistry of prominent tidal stellar streams detected in the DES\citemt{2018ApJ...862..114S}.
The details of \SSSSS\ are presented else where\citemt{Li2019a}, here we provide a summary.

\subsection{Data Reduction:}
\label{sec:data_reduction_ande_member_selection}
The streams in \SSSSS, including \phx, are identified from DES photometry\citemt{2018ApJ...862..114S}. Candidates were chosen for spectroscopic follow-up using several selections\citemt{Li2019a} that isolate the stream properties in the {\it colour-magnitude}, the {\it colour-colour} and the {\it proper motion} spaces. The observations were undertaken using the 2dF+AAOmega spectrograph on the 3.9-m Anglo-Australian Telescope (AAT) at Siding Spring Observatory in New South Wales, Australia. The 2dF field of view complements that of DECam, and its multiplexing allows for the observation of up to 392 sources across a circular, two-degree-diameter field in a single exposure.

AAOmega is a dual arm spectrograph. For these observations, the light was split into the red and blue arms by a dichroic centred at 5,800 \hbox{\AA}. Light in the blue arm was dispersed by the 580V grating; in the red arm, the 1700D was used. These gratings correspond to spectral resolutions of around 1,300 and 10,000, respectively. The respective wavelength ranges are 3,800-5,800 \hbox{\AA} in the blue arm, and 
8,400-8,820 \hbox{\AA} in the red arm, providing sufficient spectral coverage to determine kinematics and metallicity measurements from the prominent lines of the Ca\textsc{ii} triplet at about 8,600 \hbox{\AA}.
To attain sufficient signal-to-noise ratios for our faintest targets, fields were observed with a total integration time of about 7,200 seconds, typically split into three equal exposures to mitigate cosmic ray contamination.
This exposure time produces a signal-to-noise ratio of roughly 5 for targets with $r\approx18.5-19.0$, permitting us to obtain a velocity precision of roughly 1~\kms.
Seven 2dF/AAOmega fields were observed along the length of the \phx\ stream (see top panel of Extended Data Fig.~\ref{fig:gather_figure}a).
Additional calibration exposures, consisting of arc spectra and a quartz fibre flat field, were obtained with the telescope pointing at each target field, while series of bias exposures were obtained each afternoon before observing.

The data were reduced with the \textsc{2dfdr}\citeme{2015ascl.soft05015A} pipeline provided by the AAT, which automatically corrects for bias, applies a flat-field correction, calibrates the wavelength from the arc lamp exposures, traces the spectra using the flat-field exposures, and extracts the spectra. For each target the radial velocity and stellar parameters were then estimated with a dedicated pipeline\citeme{2011ApJ...736..146K} by fitting synthetic templates from the \textsc{PHOENIX} catalog grid\citeme{Husser2013A&A...553A...6H}, where the uncertainties and the means are determined by a Markov Chain Monte Carlo (MCMC) sampling of the posterior distribution. The data point with the highest signal-to-noise ratio was adopted if any particular target had multiple observations. We applied two quality criteria -- \texttt{good\_star} $= 1$ and signal-to-noise ratio $> 3$ (see ref.\citemt{Li2019a} for a definition) -- to the data to exclude obvious bad fits to the templates. 

The metallicities of the targets were derived from the equivalent widths of the Ca{\sc ii} triplet lines, using a calibration relation\citeme{Carrera2013MNRAS.434.1681C} between the equivalent widths, the absolute magnitude in the $V$ band, and the metallicity [Fe/H]. The distance\citemt{2018ApJ...862..114S} to the stream is used to calculate the absolute magnitude, hence the inferred metallicity is only valid for genuine stream members. Furthermore, the assumed calibration is applicable only to red giant branch (RGB) stars. The uncertainties on the derived metallicities are calculated from the uncertainties in the calibration relation\citeme{Carrera2013MNRAS.434.1681C} and in the equivalent width measurements\citemt{Li2019a}. Finally, the targets were cross-matched with {\it Gaia} DR2 to obtain their proper motions. 

\subsection{Member Selection:} 
\phx\ members were selected in dynamical space, based on the proper motions and the radial velocities of the targets, as follows: 
\begin{gather}
\label{eq:dynamic_select}
    -0.6\ \mathrm{mas\ yr^{-1}} < \mu_{\phi 1}  - \mu_{\phi 1,0} < 0.6\ \mathrm{mas\ yr^{-1}} \notag \\
    -0.6\ \mathrm{mas\ yr^{-1}} < \mu_{\phi 2}  - \mu_{\phi 2,0} < 0.6\ \mathrm{mas\ yr^{-1}} \notag \\
    (1.02\times\phi_{1} - 60.7)~\kms < RV_{\mathrm GSR} < (1.02\times\phi_{1} - 42.7)~\kms,
\end{gather}
Here $\phi_1$ and $\phi_2$ are the longitude and latitude, respectively, of the stream in degrees; the transformation from equitorial coordinates (right ascension, declination) to stream coordinates ($\phi_1$, $\phi_2$) uses a rotation matrix\citeme{Shipp2019}.
The proper motion in stream coordinates is ($\mu_{\phi 1}$, $\mu_{\phi 2}$), and $RV_{\mathrm GSR}$ is the line-of-sight velocity in the Galactic Standard of rest. The solar reflex motion $(11.1,240,7.3)~\kms$ (refs.~\citeme{Schonrich2010,Bland-Hawthorn2016}) was subtracted from all target proper motions and radial velocities. We assumed a distance to the Phoenix stream of $19.05\ {\rm kpc}$\citemt{2018ApJ...862..114S}, and a proper motion of the stream of $(\mu_{\phi 1,0}, \mu_{\phi 2,0}) = (-1.94, -0.36)~\rm{mas\ yr^{-1}}$ (solar reflex motion subtracted)\citeme{Shipp2019}. Proper motion gradients along the stream coordinates have previously been measured to be very low: $(d\mu_{\phi 1}/d\phi_{1}, d\mu_{\phi 2}/d\phi_{2}) = (-0.01\pm0.01,0.01\pm0.01)~\rm{mas\ yr^{-1}}$ (ref.~\citeme{Shipp2019}). Therefore, our adopted cuts in proper motion are sufficiently large to reliably include all stream members. We implemented the observed line-of-sight velocity as a third membership criterion by taking a linear cut along the stream longitude $\phi_{1}$, which is derived from a linear fitting to the proper-motion selected stars (Extended Data Fig.~\ref{fig:gather_figure}b). 

As a cross-check, we used a probabilistic mixture model to separate the Phoenix stream from a contaminating Milky Way foreground population\citeme{Li2018ApJ...866...22L, Shipp2019}.
This serves as an objective membership selection and checks the robustness of the parameter inferences in the presence of a Milky Way foreground model.
The mixture model likelihood is defined as: $\mathcal{L} = f \mathcal{L}_{\rm stream} + (1 - f) \mathcal{L}_{\rm MW}$, 
where $f$ is the fraction of stream stars and $\mathcal{L}_{\rm stream}$ and $\mathcal{L}_{\rm MW}$ are the foreground components of the stream and Milky Way. 
We consider the velocity ($RV_{\mathrm GSR}$), proper motions ($\mu_{\phi 1}$, $\mu_{\phi 2}$), and spatial position perpendicular to the stream track ($\phi_2$) of each star in our likelihood model, but exclude the spectroscopic metallicities as the Ca{\sc ii} triplet is distance dependent. Hence, the mixture model depends only on positional and dynamical information.
For velocity and proper motion likelihoods we use Gaussian distributions and include linear gradients in both velocity and proper motion space for the stream component.  We assume that the stream has no dispersion in proper motion (due to its large distance), leaving the Milky Way proper motion dispersion as a free parameter, and include a proper motion selection function based on the \SSSSS\ targeting \citemt{Li2019a}.
For the spatial likelihood, we assume a Gaussian distribution for stream stars in the $\phi_2$ direction with best-fit parameters~\citemt{2018ApJ...862..114S} and that the Milky Way foreground is constant within the stream, which weights stars closer to the stream more highly.
We compute posterior distributions using the MultiNest algorithm \citeme{Feroz2008MNRAS.384..449F, Feroz2009MNRAS.398.1601F}.
To determine stream membership, we compute the ratio of the stream to total likelihood from the posterior distribution, and take the median posterior value for each star.

We apply the mixture model to all stream targets, excluding one RR Lyrae Star, and find a total membership of 31.3 stars.
The exception is one of the BHB stars which is considered a non-member (p=0.001) due to its offset from the stream track and it is several-$\sigma$ difference from the mean Phoenix proper motion. This star has large \texttt{phot\_bp\_rp\_excess\_factor} and \texttt{astrometric\_excess\_noise\_sig} in {\it Gaia}, and there may be some unknown systematics with its proper motion.  
Additional fainter targets with larger errors are identified mostly with larger proper motions outside the selection box. The inclusion or exclusion of these stars do not change our conclusions as they do not have spectroscopic metallicities.
There are three stars that are consistent with the dynamical selection but have different position in the colour-magnitude-diagram from the other Phoenix members. Two of them are excluded because their offset in proper motion and large distance to the stream track on sky. The third star falls into the selection but its proper motion is located at the edge of the selection box with large uncertainty and it is also offset from the stream track. The equivalent-width measurement for the Ca{\sc ii} triplet of this star, assuming it is the member of \phx, yields a much higher $\feh = -0.48 \pm 0.24$, which significantly deviates from the metallicity of the stream. Hence we do not consider any of these as members of the \phx\ stream.

In summary, we identified 25 member stars in the Phoenix stream with robust RV measurements, including three BHBs and one RRLyrae; among them, 11 RGB members have signal-to-noise ratio greater than 10; their Ca{\sc ii} triplet metallicities are used for our analysis.
Extended Data Fig.~\ref{fig:gather_figure}a shows the on-sky distribution of these stars, colour-coded by their metallicity, demonstrating the physical narrowness of the \phx\ stream. Also shown are the other stars targeted within each of the 2dF fields. 
Extended Data Fig.~\ref{fig:gather_figure}c shows the resultant colour-magnitude diagram of the \phx\ members in
($ g_{\mathrm DECam}-r_{\mathrm DECam},g_{\mathrm DECam}$), as well as 10 \textsc{padova} isochrones~\citeme{2017ApJ...835...77M} with an age of $11.2\ \mathrm{Gyr}$ and metallicities spanning $\feh = -2.0$ to $\feh = -2.9$. The isochrones reproduce the stellar sequence well---including the main sequence turnoff, the RGB and the horizontal branch. We excluded the BHB and RR Lyrae stars (marked with orange squares in Extended Data Fig.~\ref{fig:gather_figure}) from our metallicity analysis, as the Ca{\sc ii} triplet metallicity calibration only applies to RGB stars.

\subsection{Stream Metallicity and Radial Velocity:}
To determine the mean and intrinsic width of the metallicity distribution of the \phx\ stream, we focus upon the 11 RGB stars identified as member and with spectroscopic signal-to-noise ration greater than 10 (as shown in Fig.~\ref{fig:distmetal}). We represent the metallicity $\feh$ distribution as a Gaussian of the form 
\begin{equation}
    P( \feh_i ) = \frac{1}{\sqrt{2 \pi \sigma_{\feh}^2}} \exp{ \left(  - \frac{( \feh_i - \feh )^2}{2 \sigma_{\feh}^2 } \right) }  
\end{equation}
where $\feh$ is the mean metallicity and $\sigma_{\feh}$ is the intrinsic width. We use a MCMC approach to explore the likelihood space, which was calculated by convolving the above distribution with the individual metallicity uncertainties. The resultant posterior distributions are presented in the corner plot in Extended Data Fig.~\ref{fig:feh_corner}. From the marginalised posterior distributions, we infer $\feh = -2.70\pm{0.06}$ and an intrinsic dispersion $\sigma_{\feh} < 0.2$ at 95\% confidence.
We aim to present further analyses on the spread of other individual elements, such as sodium, with follow-up observations of the \phx\ Stream.

Extended Data Fig.~\ref{fig:gather_figure}b shows the radial velocity in the Galactic Standard of rest $RV_{\mathrm GSR}$ for all stars selected using our proper motion cuts. Members of \phx\ stream are shown as circles between the dashed lines.
It is clear that $RV_{\mathrm GSR}$ peaks at around -50~\kms. 
We represent the velocity $v$ distribution as a Gaussian of the form
\begin{equation}
    P(v_i) = \frac{1}{\sqrt{2 \pi \sigma_{\mathrm RV}^2 }} \exp{ \left(  
    -\frac{( v_i - RV_{\mathrm GSR}( \phi_1 ) )^2 }{ 2 \sigma_{\mathrm RV}^2 }
    \right) }
\end{equation}
where $\sigma_{\mathrm RV}$ is the velocity dispersion of the stream and $RV_{\mathrm GSR}( \phi_1 ) = p_0 + p_1\phi_1 + p_2\phi_1^2$ is a second-order polynomial fitted to the stream velocity as a function of stream longitude ($\phi_1$). We again use an MCMC approach to explore the joint likelihood space for the polynomial coefficients and the velocity dispersion, $\sigma_{\mathrm RV}$, with the likelihood formed by convolving the above distribution with the individual velocity errors. The resultant posterior distributions are presented in Extended Data Fig.~\ref{fig:vgsr_corner}; we infer the intrinsic velocity dispersion to be $\sigma_{\mathrm RV} = 2.66^{+0.71}_{-0.57}$~\kms\ from its marginalised posterior distribution.
We note the intrinsic velocity dispersion of the progenitor globular cluster could have be larger than the stream velocity dispersion owing to the specific details of how stars are tidally stripped. 

\subsection{Dynamical Modelling:}
We modelled the dynamics of the \phx\ stream using established numerical techniques \citeme{orphan_lmc_modelling,gibbons_etal_2014}, considering the Milky Way and the gravitational influence of its largest satellite, the LMC. For the Milky Way, we used a best-fit potential\citeme{mcmillan_2017}, in particular, we use the implementation of the potential from \texttt{galpot} \citeme{dehnen_binney_1998}. The LMC is modelled as an Hernquist profile \citeme{hernquist_1990} with a mass of $1.5\times10^{11} M_\odot$ and a scale radius of $17.14$ kpc, consistent with the recent measurement of the LMC mass\citeme{orphan_lmc_modelling}. During the fit, we keep the potential fixed and vary only the present-day proper motions, radial velocity, distance and on-sky location of \phx's progenitor. For simplicity, we place the progenitor at a location of $\phi_1 = 0^\circ$. We fit all of the high-likelihood members from this work, taking into account their position on the sky, proper motions and radial velocities. For the distance, we use a prior\citemt{2018ApJ...862..114S} of $19.1\pm1.0$ kpc. The stream is evolved for 3 Gyr which is more than sufficient to cover the observed portion of Phoenix. The progenitor is modelled as a Plummer sphere with a mass of $2\times10^4 M_\odot$ and a scale radius of 10 pc. Because the progenitor of \phx\ has not been located within the observed portion of the \phx, the mass of the progenitor is linearly interpolated from its initial value to zero at the present-day. We use an MCMC implementation~\citeme{emcee} to explore the posterior space, with 100 walkers for 1,000 steps and a burn-in of 500 steps. To account for our uncertainty in the potential, we repeat this procedure nine additional times using potential parameters drawn from the posterior distribution of the fits~\citeme{mcmillan_2017}. 

Our best fit orbit reproduces the key data for Phoenix (Extended Data Fig.~\ref{fig:sim}). We find that the stream orbits in a prograde direction with an inclination of roughly $60^\circ$ relative to the Milky Way disk. Because of this, \phx\ will be sensitive to baryonic substructure in the Milky Way disk \citeme{amorisco_gmcs,erkal_etal_2017,pearson_pal5,banik_bovy_2019}. The inferred orbit has a pericenter of $12.9^{+0.3}_{-0.5}$ kpc, an apocenter of $18.4^{+0.3}_{-0.2}$ kpc, and an eccentricity of $0.18\pm{0.01}$. The best fit model places the stream at a distance of approximately 17.5 kpc, slightly closer than estimated from isochrone fitting. If the stream is located at this closer distance, this would increase the metallicity by only about 0.04, which does not affect our conclusions. We find that it takes about 2 Gyr of tidal disruption to produce the observed length of the Phoenix stream.

To compare the dynamical properties of \phx\ to those for the population of Milky Way globular clusters, in particular, the integral of motion commonly referred to as the `action', we use \textsc{agama} \citeme{agama}. For \phx, we use the posterior chains of the MCMC fits (done in the best-fit potential from\citeme{mcmillan_2017}) to compute its mean actions and energy. For the globular clusters, we Monte Carlo sample each globular cluster's present-day phase-space position 50 times to get the uncertainty in the actions\citemt{vasiliev_2019}. The results are in Extended Data Fig.\ref{fig:actions}. Note that we have updated the distance to Pal 5\citeme{2019AJ....158..223P}. Interestingly, this cluster, followed by NGC5053, is the closest to \phx\ in energy and action space. However, the orbital plane has a significantly different azimuthal orientation (about 80$^\circ$) from Pal 5, so these streams are not directly connected but may have been accreted together. We also explore the potential connection between the Phoenix and Hermus stream\citemt{2016ApJ...820L..27G}. We find that the continuation of our stream model passes through the location of Hermus on the sky. Furthermore, our best-fit \phx\ model matches the orbital inclination of Hermus \citemt{2016ApJ...820L..27G}. We find that it requires around 8 Gyr of disruption to produce a stream long enough to connect Phoenix and Hermus which indicates that they are not directly connected, but may have been accreted with the same dwarf galaxy progenitor. We also perform fits without the LMC and find that Phoenix can be accurately fit in either case. Thus, Phoenix is not as sensitive to the LMC as is the Orphan stream\citeme{orphan_lmc_modelling,2019MNRAS.485.4726K}.

\end{methods}

\subsection{Data Availability}
The data used in this paper is from the \SSSSS internal data release version 1.5; see \url{https://s5collab.github.io}. The first public data release is scheduled on the end of 2020, which contains the observation taken in 2018 and 2019.
Data is available on request from
 Ting Li (tingli@carnegiescience.edu). Source data are provided with this paper.

\subsection{Code Availability}
The \textsc{2dfdr} for the raw data reduction is available at \url{https://www.aao.gov.au/science/software/2dfdr}.
The \textsc{rvspecfit} \citeme{2011ApJ...736..146K} used for the determination of stellar parameters is available from \url{https://github.com/segasai/rvspecfit}. Other analysis code will be made available on request.

\renewcommand{\figurename}{Extended Data Figure}
\setcounter{figure}{0}

\begin{figure}
    \centering
    \includegraphics[width=0.54\linewidth]{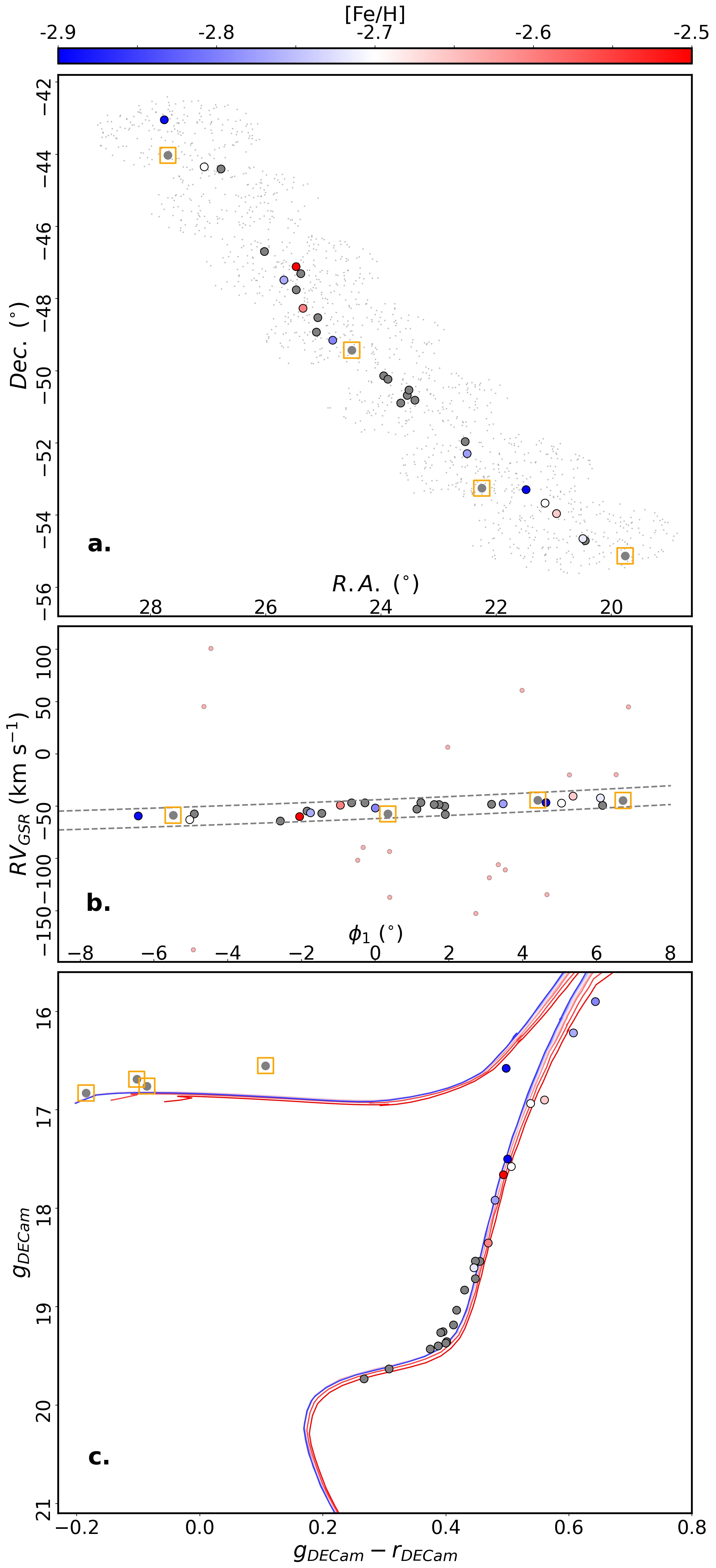}
\end{figure}

\begin{figure}
    \centering
    \includegraphics[width=0.0\linewidth]{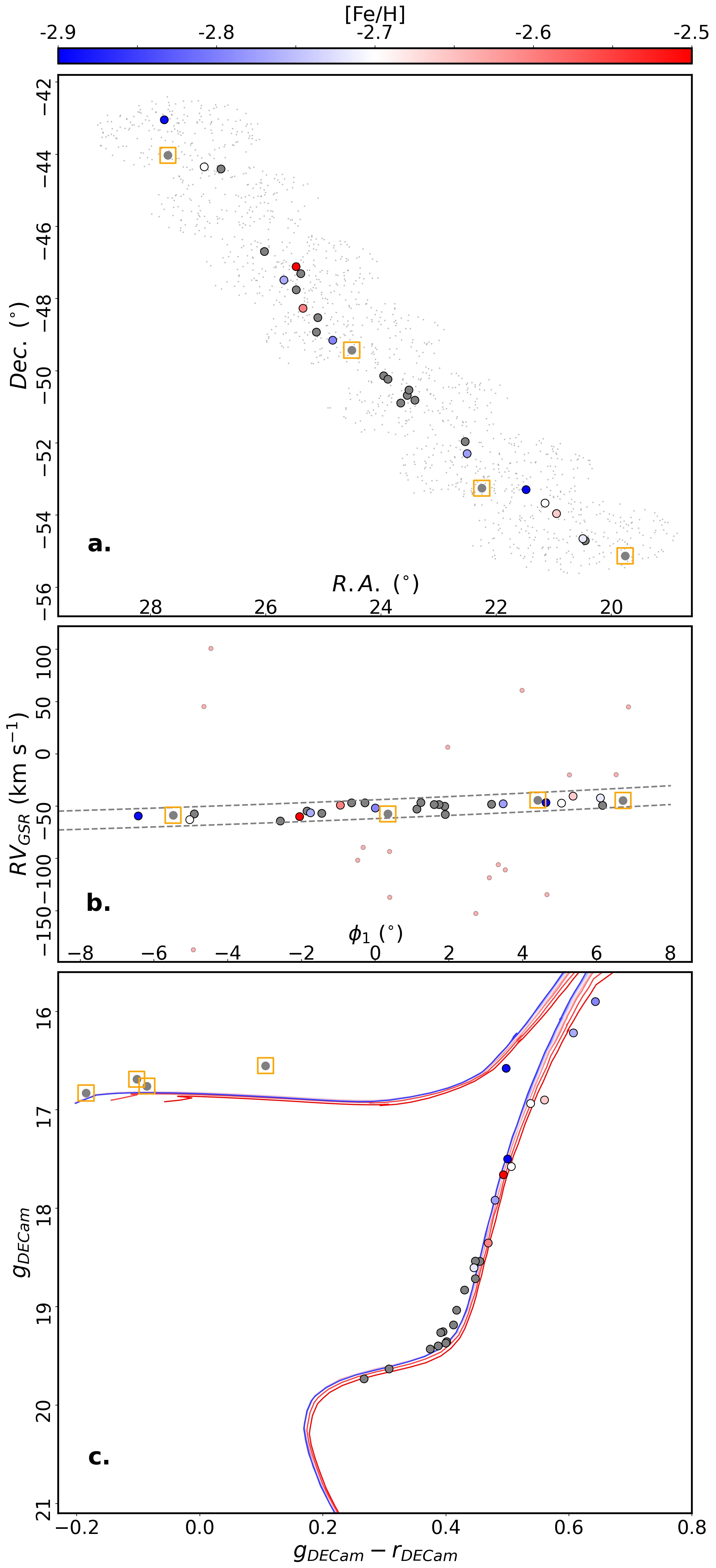}
    \caption{{\it a}: The on-sky distribution of all stars observed in the 2dF fields targeting the Phoenix Stream. The overall footprint is a series of circular 2dF pointings. R.A., right ascension; Dec., declination.  {\it b}: Radial velocity in the Galactic Standard of rest ($RV_{\mathrm GSR}$) versus stream longitude ($\phi_1$) for \phx\ stars selected on the basis of proper motion, photometry and the mixture model. On the basis of the approximately linear correlation between $RV_{\mathrm GSR}$ and $\phi_1$, we select \phx\ stream members from the region between the dashed lines as Eq 1 describes, which effectively excludes non-members (shown as small pink circles). {\it c}: The colour-magnitude diagram of members of the \phx\ stream.
    Over-plotted are \textsc{padova} isochrones  \protect\citeme{2017ApJ...835...77M} with $\feh = -2.9$ to $\feh = -2.0$ (from blue to red), $m - M = 16.4$(ref.\protect\citetmt{2018ApJ...862..114S}), where $m - M$ is the distance modulus,
    $m$ is the apparent magnitude and $M$ is the absolute magnitude) and $\mathrm{log}_{10}(\mathrm{age/Gyr}) = 10.05$. 
    In all panels, the stars we identify as members of the \phx\ stream are represented by large circles; those with high signal-to-noise ratio are colour-coded by their metallicity, others are gray. The four orange squares indicate the BHB/RR Lyrae stars, metallicities of which cannot be measured with the method used here.}
    \label{fig:gather_figure}
\end{figure}

\begin{figure}
    \centering
    \includegraphics[width=0.7\linewidth]{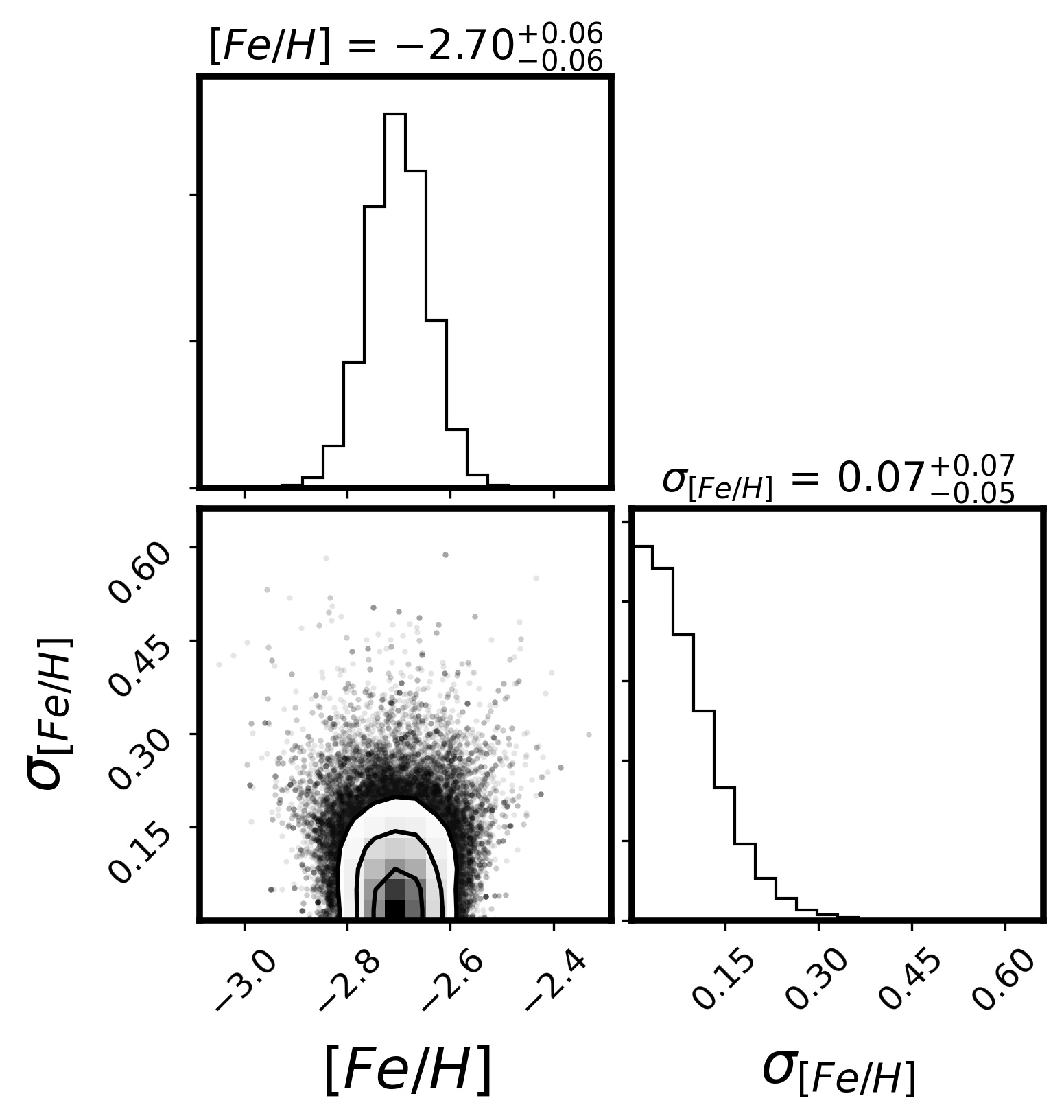}
    \caption{The posterior sampling results of the metallicity distribution of the 11 Phoenix member stars with signal-to-noise ratios greater than 10. The mean and dispersion of the metallicity are noted. The dispersion is consistent with being zero, with $\sigma_{\feh} < 0.2$ being the 95\% confidence interval. This figure is made using the corner package \protect\citeme{corner}. }
    \label{fig:feh_corner}
\end{figure}

\begin{figure}
    \centering
    \includegraphics[width=0.7\linewidth]{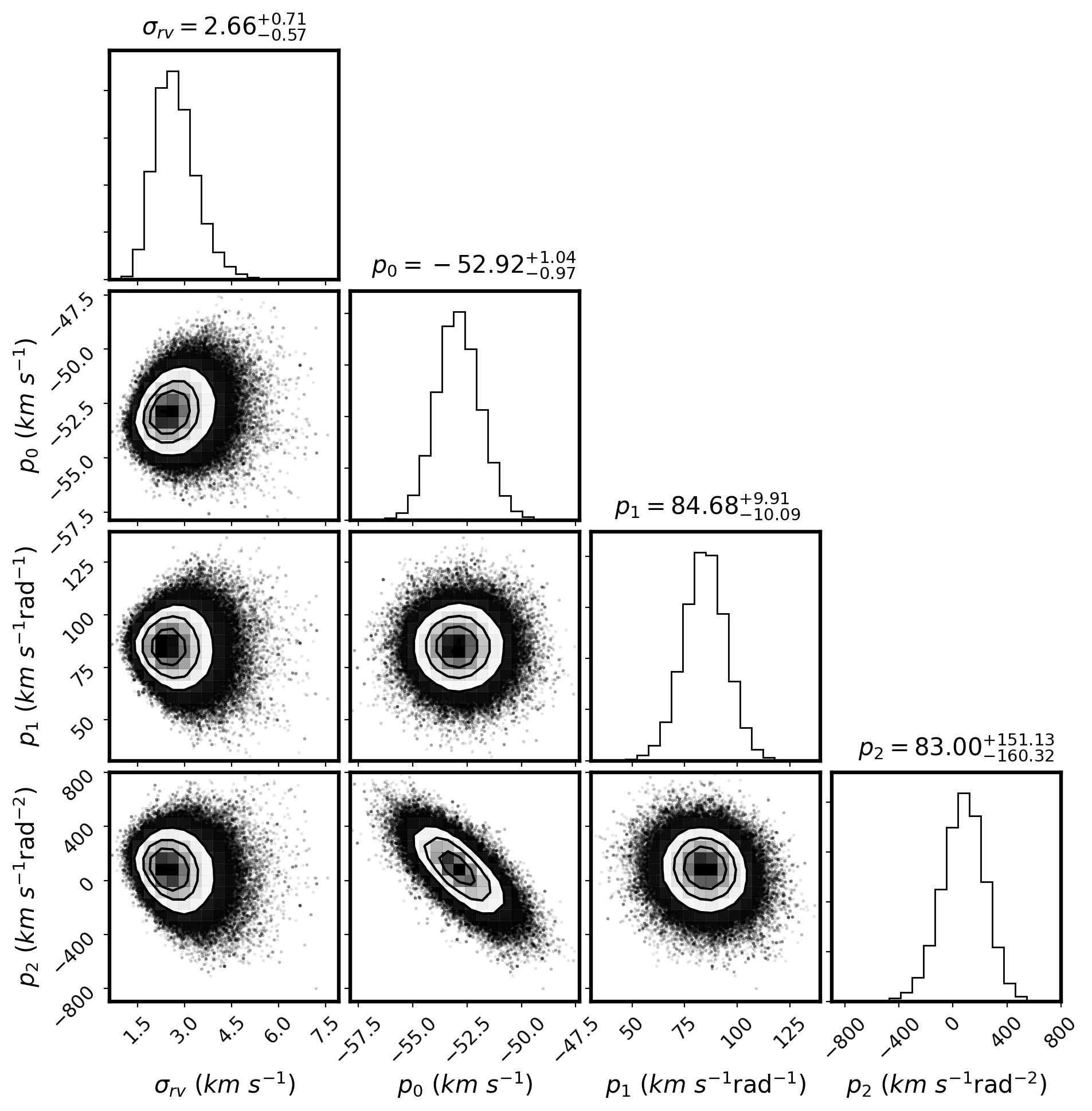}
    \caption{The posterior sampling results of the RV$_{GSR}$ distribution. The parameters $p_0$,$p_1$, and $p_2$, are the best-fitting polynomial parameters for $RV_{GSR} (\phi_1) = p_0 + p_1\phi_1 + p_2\phi_1^{2}$; $\sigma_{rv}$ is the intrinsic dispersion. Here the best-fitting parameters are calculated with $phi_1$ in radians. This figure is made with {\it corner} package \protect\citeme{corner}}
    \label{fig:vgsr_corner}
\end{figure}

\begin{figure}
    \centering
    \includegraphics[width=0.8\linewidth]{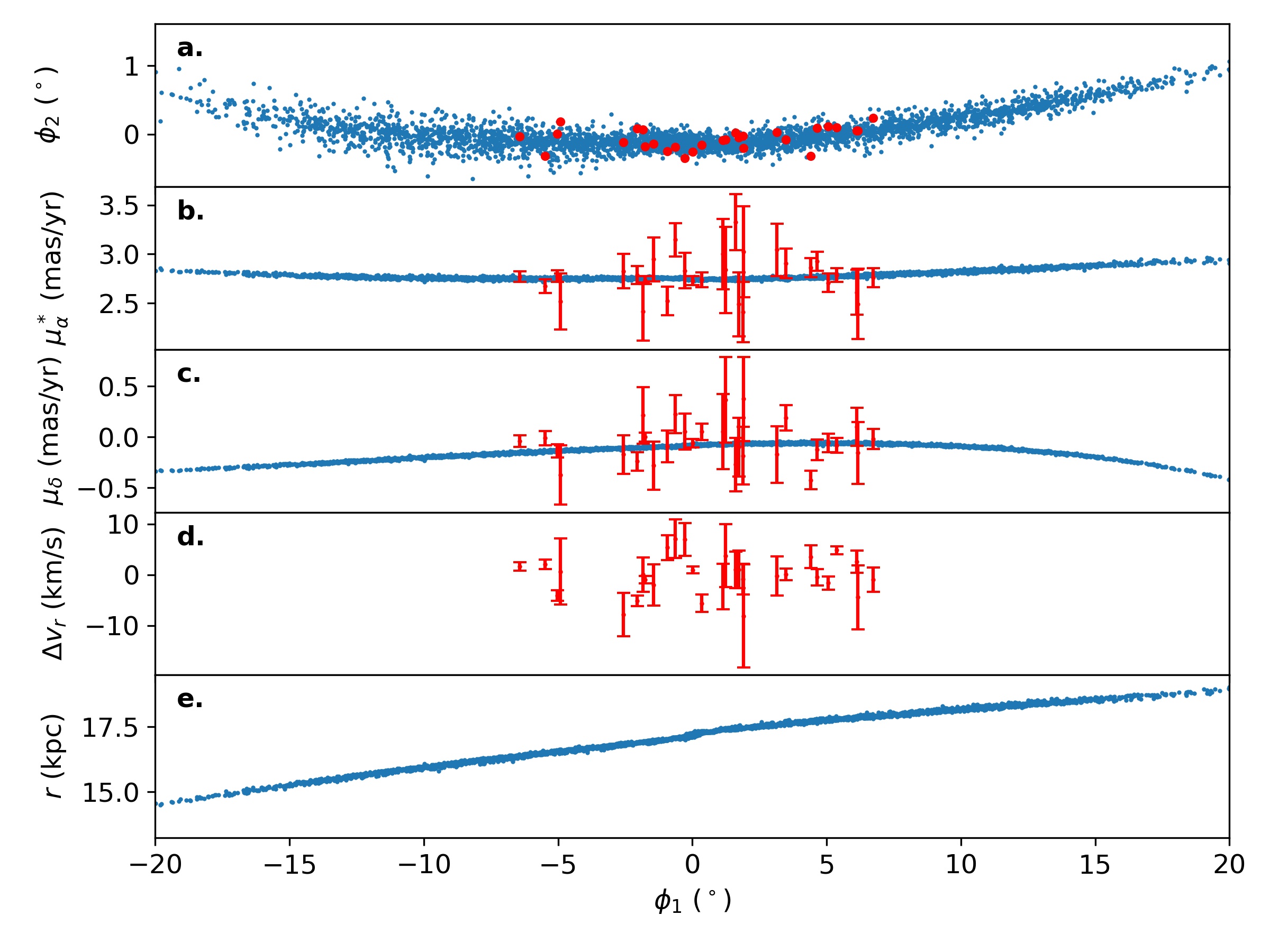}
    \caption{Best-fit model to the Phoenix stream. a-e, The stream on the sky (a), the proper motions in the right ascension ($\mu_{\alpha}^{\*}$; b) and declination ($\mu_{\delta}$; c), the residuals of the radial velocity ($\Delta v_r$; d) and the distance to the stream ($r$, e).The blue points show the best-fit model and the red points (a) or error bars (b-d; 1$\sigma$ uncertainty) show the observed values.}
    \label{fig:sim}
\end{figure}

\begin{figure}
    \centering
    \includegraphics[width=0.99\linewidth]{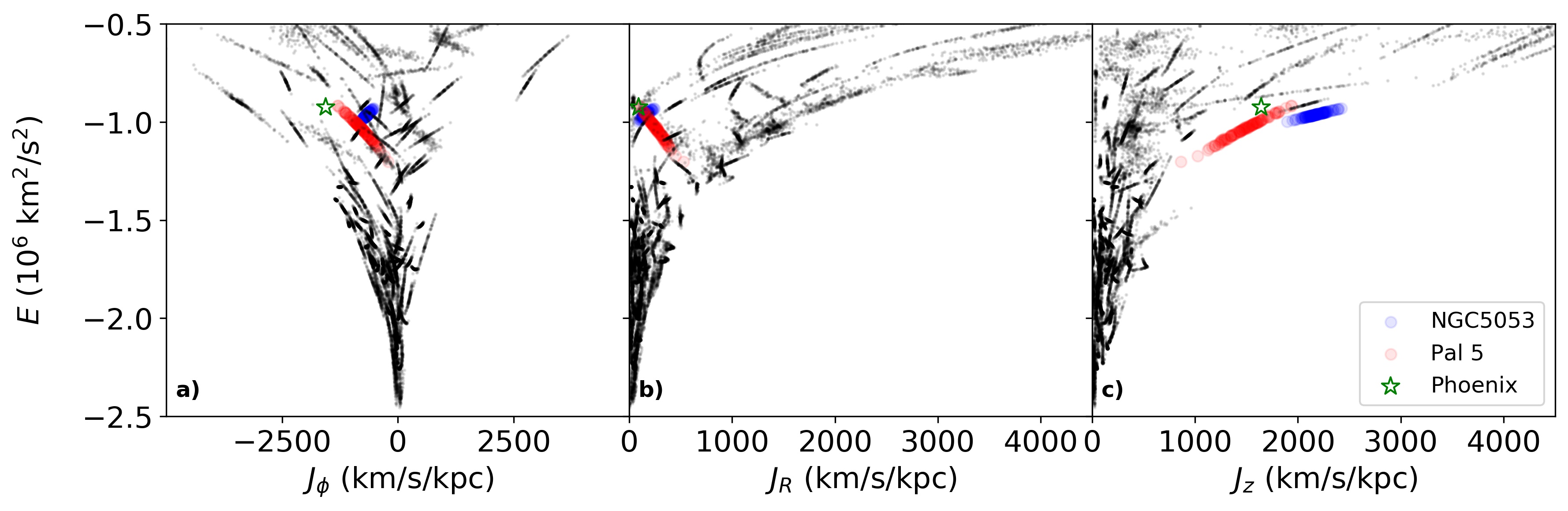}
    \caption{Comparison of energy $E$ and actions $J_{\phi, R, z}$ for the Phoenix stream and all Milky Way globular clusters.a-c, The actions are computed with \textsc{agama}\protect\citeme{agama} in the best-fit Milky Way potential\protect\citeme{mcmillan_2017}. Pal 5 (red circles) is closest in energy and actions to the Phoenix stream (green star), suggesting a possible association. There is also a potential relation in this space to the NGC5053 (blue circles), another globular cluster. All other globular clusters are shown in black.} 
    \label{fig:actions}
\end{figure}

\clearpage


\begin{thebibliography}{10}
\expandafter\ifx\csname url\endcsname\relax
  \def\url#1{\texttt{#1}}\fi
\expandafter\ifx\csname urlprefix\endcsname\relax\def\urlprefix{URL }\fi
\providecommand{\bibinfo}[2]{#2}
\providecommand{\eprint}[2][]{\url{#2}}

\bibitem{1991ARA&A..29..543H}
\bibinfo{author}{{Harris}, W.~E.}
\newblock \bibinfo{title}{{Globular cluster systems in galaxies beyond the
  Local Group.}}
\newblock \emph{\bibinfo{journal}{Annual Reviews of Astronomy \& Astrophysics}}
  \textbf{\bibinfo{volume}{29}}, \bibinfo{pages}{543--579}
  (\bibinfo{year}{1991}).

\bibitem{2006ARA&A..44..193B}
\bibinfo{author}{{Brodie}, J.~P.} \& \bibinfo{author}{{Strader}, J.}
\newblock \bibinfo{title}{{Extragalactic Globular Clusters and Galaxy
  Formation}}.
\newblock \emph{\bibinfo{journal}{Annual Reviews of Astronomy \& Astrophysics}}
  \textbf{\bibinfo{volume}{44}}, \bibinfo{pages}{193--267}
  (\bibinfo{year}{2006}).
\newblock \eprint{astro-ph/0602601}.

\bibitem{2019Natur.574...69M}
\bibinfo{author}{{Mackey}, D.} \emph{et~al.}
\newblock \bibinfo{title}{{Two major accretion epochs in M31 from two distinct
  populations of globular clusters}}.
\newblock \emph{\bibinfo{journal}{\nat}} \textbf{\bibinfo{volume}{574}},
  \bibinfo{pages}{69--71} (\bibinfo{year}{2019}).
\newblock \eprint{1910.00808}.

\bibitem{1996AJ....112.1487H}
\bibinfo{author}{{Harris}, W.~E.}
\newblock \bibinfo{title}{{A Catalog of Parameters for Globular Clusters in the
  Milky Way}}.
\newblock \emph{\bibinfo{journal}{\aj}} \textbf{\bibinfo{volume}{112}},
  \bibinfo{pages}{1487} (\bibinfo{year}{1996}).

\bibitem{2018RSPSA.47470616F}
\bibinfo{author}{{Forbes}, D.~A.} \emph{et~al.}
\newblock \bibinfo{title}{{Globular cluster formation and evolution in the
  context of cosmological galaxy assembly: open questions}}.
\newblock \emph{\bibinfo{journal}{Proceedings of the Royal Society of London
  Series A}} \textbf{\bibinfo{volume}{474}}, \bibinfo{pages}{20170616}
  (\bibinfo{year}{2018}).
\newblock \eprint{1801.05818}.

\bibitem{2019MNRAS.487.1986B}
\bibinfo{author}{{Beasley}, M.~A.} \emph{et~al.}
\newblock \bibinfo{title}{{An old, metal-poor globular cluster in Sextans A and
  the metallicity floor of globular cluster systems}}.
\newblock \emph{\bibinfo{journal}{\mnras}} \textbf{\bibinfo{volume}{487}},
  \bibinfo{pages}{1986--1993} (\bibinfo{year}{2019}).
\newblock \eprint{1904.01084}.

\bibitem{2019MNRAS.486L..20K}
\bibinfo{author}{{Kruijssen}, J.~M.~D.}
\newblock \bibinfo{title}{{The minimum metallicity of globular clusters and its
  physical origin - implications for the galaxy mass-metallicity relation and
  observations of proto-globular clusters at high redshift}}.
\newblock \emph{\bibinfo{journal}{\mnras}} \textbf{\bibinfo{volume}{486}},
  \bibinfo{pages}{L20--L25} (\bibinfo{year}{2019}).
\newblock \eprint{1904.09987}.

\bibitem{2016ApJ...820...58B}
\bibinfo{author}{{Balbinot}, E.} \emph{et~al.}
\newblock \bibinfo{title}{{The Phoenix Stream: A Cold Stream in the Southern
  Hemisphere}}.
\newblock \emph{\bibinfo{journal}{\apj}} \textbf{\bibinfo{volume}{820}},
  \bibinfo{pages}{58} (\bibinfo{year}{2016}).
\newblock \eprint{1509.04283}.

\bibitem{2018ApJ...862..114S}
\bibinfo{author}{{Shipp}, N.} \emph{et~al.}
\newblock \bibinfo{title}{{Stellar Streams Discovered in the Dark Energy
  Survey}}.
\newblock \emph{\bibinfo{journal}{\apj}} \textbf{\bibinfo{volume}{862}},
  \bibinfo{pages}{114} (\bibinfo{year}{2018}).
\newblock \eprint{1801.03097}.

\bibitem{Erkal:2016}
\bibinfo{author}{{Erkal}, D.}, \bibinfo{author}{{Sanders}, J.~L.} \&
  \bibinfo{author}{{Belokurov}, V.}
\newblock \bibinfo{title}{{Stray, swing and scatter: angular momentum evolution
  of orbits and streams in aspherical potentials}}.
\newblock \emph{\bibinfo{journal}{\mnras}} \textbf{\bibinfo{volume}{461}},
  \bibinfo{pages}{1590--1604} (\bibinfo{year}{2016}).
\newblock \eprint{1603.08922}.

\bibitem{2016ApJ...820L..27G}
\bibinfo{author}{{Grillmair}, C.~J.} \& \bibinfo{author}{{Carlberg}, R.~G.}
\newblock \bibinfo{title}{{What a Tangled Web We Weave: Hermus as the Northern
  Extension of the Phoenix Stream}}.
\newblock \emph{\bibinfo{journal}{\apjl}} \textbf{\bibinfo{volume}{820}},
  \bibinfo{pages}{L27} (\bibinfo{year}{2016}).
\newblock \eprint{1603.02278}.

\bibitem{2016ApJ...830..135C}
\bibinfo{author}{{Carlberg}, R.~G.} \& \bibinfo{author}{{Grillmair}, C.~J.}
\newblock \bibinfo{title}{{Velocity Variations in the Phoenix-Hermus Star
  Stream}}.
\newblock \emph{\bibinfo{journal}{\apj}} \textbf{\bibinfo{volume}{830}},
  \bibinfo{pages}{135} (\bibinfo{year}{2016}).
\newblock \eprint{1606.05769}.

\bibitem{Li2019a}
\bibinfo{author}{Li, T.~S.} \emph{et~al.}
\newblock \bibinfo{title}{{The Southern Stellar Stream Spectroscopic Survey
  (S5): Overview, Target Selection, Data Reduction, Validation, and Early
  Science}}.
\newblock \emph{\bibinfo{journal}{\mnras}}  (\bibinfo{year}{2019}).
\newblock \eprint{1907.09481}.

\bibitem{Abbott2018}
\bibinfo{author}{{Abbott}, T.~M.~C.} \emph{et~al.}
\newblock \bibinfo{title}{{The Dark Energy Survey: Data Release 1}}.
\newblock \emph{\bibinfo{journal}{\apjs}} \textbf{\bibinfo{volume}{239}},
  \bibinfo{pages}{18} (\bibinfo{year}{2018}).
\newblock \eprint{1801.03181}.

\bibitem{2016A&A...595A...1G}
\bibinfo{author}{{Gaia Collaboration}} \emph{et~al.}
\newblock \bibinfo{title}{{The Gaia mission}}.
\newblock \emph{\bibinfo{journal}{\aap}} \textbf{\bibinfo{volume}{595}},
  \bibinfo{pages}{A1} (\bibinfo{year}{2016}).
\newblock \eprint{1609.04153}.

\bibitem{2018A&A...616A...1G}
\bibinfo{author}{{Gaia Collaboration}} \emph{et~al.}
\newblock \bibinfo{title}{{Gaia Data Release 2. Summary of the contents and
  survey properties}}.
\newblock \emph{\bibinfo{journal}{\aap}} \textbf{\bibinfo{volume}{616}},
  \bibinfo{pages}{A1} (\bibinfo{year}{2018}).
\newblock \eprint{1804.09365}.

\bibitem{2019MNRAS.482.1275U}
\bibinfo{author}{{Usher}, C.} \emph{et~al.}
\newblock \bibinfo{title}{{The WAGGS project - II. The reliability of the
  calcium triplet as a metallicity indicator in integrated stellar light}}.
\newblock \emph{\bibinfo{journal}{\mnras}} \textbf{\bibinfo{volume}{482}},
  \bibinfo{pages}{1275--1303} (\bibinfo{year}{2019}).
\newblock \eprint{1809.07650}.

\bibitem{Simon2019ARA&A..57..375S}
\bibinfo{author}{{Simon}, J.~D.}
\newblock \bibinfo{title}{{The Faintest Dwarf Galaxies}}.
\newblock \emph{\bibinfo{journal}{\araa}} \textbf{\bibinfo{volume}{57}},
  \bibinfo{pages}{375--415} (\bibinfo{year}{2019}).
\newblock \eprint{1901.05465}.

\bibitem{Willman2012AJ....144...76W}
\bibinfo{author}{{Willman}, B.} \& \bibinfo{author}{{Strader}, J.}
\newblock \bibinfo{title}{{``Galaxy,'' Defined}}.
\newblock \emph{\bibinfo{journal}{\aj}} \textbf{\bibinfo{volume}{144}},
  \bibinfo{pages}{76} (\bibinfo{year}{2012}).
\newblock \eprint{1203.2608}.

\bibitem{Simon2018ApJ...863...89S}
\bibinfo{author}{{Simon}, J.~D.}
\newblock \bibinfo{title}{{Gaia Proper Motions and Orbits of the Ultra-faint
  Milky Way Satellites}}.
\newblock \emph{\bibinfo{journal}{\apj}} \textbf{\bibinfo{volume}{863}},
  \bibinfo{pages}{89} (\bibinfo{year}{2018}).
\newblock \eprint{1804.10230}.

\bibitem{Helmi2018A&A...616A..12G}
\bibinfo{author}{{Gaia Collaboration}} \emph{et~al.}
\newblock \bibinfo{title}{{Gaia Data Release 2. Kinematics of globular clusters
  and dwarf galaxies around the Milky Way}}.
\newblock \emph{\bibinfo{journal}{\aap}} \textbf{\bibinfo{volume}{616}},
  \bibinfo{pages}{A12} (\bibinfo{year}{2018}).
\newblock \eprint{1804.09381}.

\bibitem{vasiliev_2019}
\bibinfo{author}{{Vasiliev}, E.}
\newblock \bibinfo{title}{{Proper motions and dynamics of the Milky Way
  globular cluster system from Gaia DR2}}.
\newblock \emph{\bibinfo{journal}{\mnras}} \textbf{\bibinfo{volume}{484}},
  \bibinfo{pages}{2832--2850} (\bibinfo{year}{2019}).
\newblock \eprint{1807.09775}.

\bibitem{Simpson:2018cm}
\bibinfo{author}{Simpson, J.~D.}
\newblock \bibinfo{title}{{The most metal-poor Galactic globular cluster: the
  first spectroscopic observations of ESO280-SC06}}.
\newblock \emph{\bibinfo{journal}{\mnras}} \textbf{\bibinfo{volume}{477}},
  \bibinfo{pages}{4565--4576} (\bibinfo{year}{2018}).

\bibitem{Simpson2019a}
\bibinfo{author}{Simpson, J.~D.} \& \bibinfo{author}{Martell, S.~L.}
\newblock \bibinfo{title}{{A Nitrogen-Enhanced Metal-Poor star discovered in
  the globular cluster ESO280-SC06}}.
\newblock \emph{\bibinfo{journal}{\mnras}} \textbf{\bibinfo{volume}{490}},
  \bibinfo{pages}{741--751} (\bibinfo{year}{2019}).

\bibitem{Larsen:2012cb}
\bibinfo{author}{Larsen, S.~S.}, \bibinfo{author}{Brodie, J.~P.} \&
  \bibinfo{author}{Strader, J.}
\newblock \bibinfo{title}{{Detailed abundance analysis from integrated
  high-dispersion spectroscopy: globular clusters in the Fornax dwarf
  spheroidal}}.
\newblock \emph{\bibinfo{journal}{\aap}} \textbf{\bibinfo{volume}{546}},
  \bibinfo{pages}{A53} (\bibinfo{year}{2012}).

\bibitem{Kruijssen:2015fe}
\bibinfo{author}{Kruijssen, J. M.~D.}
\newblock \bibinfo{title}{{Globular clusters as the relics of regular star
  formation in ‘normal' high-redshift galaxies}}.
\newblock \emph{\bibinfo{journal}{\mnras}} \textbf{\bibinfo{volume}{454}},
  \bibinfo{pages}{1658--1686} (\bibinfo{year}{2015}).

\bibitem{2019MNRAS.482.3868I}
\bibinfo{author}{{Iorio}, G.} \& \bibinfo{author}{{Belokurov}, V.}
\newblock \bibinfo{title}{{The shape of the Galactic halo with Gaia DR2 RR
  Lyrae. Anatomy of an ancient major merger}}.
\newblock \emph{\bibinfo{journal}{\mnras}} \textbf{\bibinfo{volume}{482}},
  \bibinfo{pages}{3868--3879} (\bibinfo{year}{2019}).
\newblock \eprint{1808.04370}.

\bibitem{Roederer2019}
\bibinfo{author}{{Roederer}, I.~U.} \& \bibinfo{author}{{Gnedin}, O.~Y.}
\newblock \bibinfo{title}{{High-resolution Optical Spectroscopy of Stars in the
  Sylgr Stellar Stream}}.
\newblock \emph{\bibinfo{journal}{\apj}} \textbf{\bibinfo{volume}{883}},
  \bibinfo{pages}{84} (\bibinfo{year}{2019}).
\newblock \eprint{1907.03772}.

\bibitem{2017MNRAS.469L..63R}
\bibinfo{author}{{Renzini}, A.}
\newblock \bibinfo{title}{{Finding forming globular clusters at high
  redshifts}}.
\newblock \emph{\bibinfo{journal}{\mnras}} \textbf{\bibinfo{volume}{469}},
  \bibinfo{pages}{L63--L67} (\bibinfo{year}{2017}).
\newblock \eprint{1704.04883}.

\bibitem{SimpsonJ2020}
\bibinfo{author}{{Simpson}, J.~D.}
\newblock \bibinfo{title}{{mpirical relationship between calcium triplet
  equivalent widths and [Fe/H] using Gaia photometry (version 0.2) [data
  set].}}
\newblock \emph{\bibinfo{journal}{Zenodo}}  (\bibinfo{year}{2020}).

\end{thebibliography}

\begin{thebibliography}{10}
\expandafter\ifx\csname url\endcsname\relax
  \def\url#1{\texttt{#1}}\fi
\expandafter\ifx\csname urlprefix\endcsname\relax\def\urlprefix{URL }\fi
\providecommand{\bibinfo}[2]{#2}
\providecommand{\eprint}[2][]{\url{#2}}

\bibitem{2015ascl.soft05015A}
\bibinfo{author}{{AAO Software Team}}.
\newblock \bibinfo{title}{{2dfdr: Data reduction software}}
  (\bibinfo{year}{2015}).
\newblock \eprint{1505.015}.

\bibitem{2011ApJ...736..146K}
\bibinfo{author}{{Koposov}, S.~E.} \emph{et~al.}
\newblock \bibinfo{title}{{Accurate Stellar Kinematics at Faint Magnitudes:
  Application to the Bo{\"o}tes I Dwarf Spheroidal Galaxy}}.
\newblock \emph{\bibinfo{journal}{\apj}} \textbf{\bibinfo{volume}{736}},
  \bibinfo{pages}{146} (\bibinfo{year}{2011}).
\newblock \eprint{1105.4102}.

\bibitem{Husser2013A&A...553A...6H}
\bibinfo{author}{{Husser}, T.~O.} \emph{et~al.}
\newblock \bibinfo{title}{{A new extensive library of PHOENIX stellar
  atmospheres and synthetic spectra}}.
\newblock \emph{\bibinfo{journal}{\aap}} \textbf{\bibinfo{volume}{553}},
  \bibinfo{pages}{A6} (\bibinfo{year}{2013}).
\newblock \eprint{1303.5632}.

\bibitem{Carrera2013MNRAS.434.1681C}
\bibinfo{author}{{Carrera}, R.}, \bibinfo{author}{{Pancino}, E.},
  \bibinfo{author}{{Gallart}, C.} \& \bibinfo{author}{{del Pino}, A.}
\newblock \bibinfo{title}{{The near-infrared Ca II triplet as a metallicity
  indicator - II. Extension to extremely metal-poor metallicity regimes}}.
\newblock \emph{\bibinfo{journal}{\mnras}} \textbf{\bibinfo{volume}{434}},
  \bibinfo{pages}{1681--1691} (\bibinfo{year}{2013}).
\newblock \eprint{1306.3883}.

\bibitem{Shipp2019}
\bibinfo{author}{Shipp, N.} \emph{et~al.}
\newblock \bibinfo{title}{{Proper Motions of Stellar Streams Discovered in the
  Dark Energy Survey}}.
\newblock \emph{\bibinfo{journal}{\apj}} \textbf{\bibinfo{volume}{885}},
  \bibinfo{pages}{3} (\bibinfo{year}{2019}).
\newblock \eprint{1907.09488}.

\bibitem{Schonrich2010}
\bibinfo{author}{{Sch{\"o}nrich}, R.}, \bibinfo{author}{{Binney}, J.} \&
  \bibinfo{author}{{Dehnen}, W.}
\newblock \bibinfo{title}{{Local kinematics and the local standard of rest}}.
\newblock \emph{\bibinfo{journal}{\mnras}} \textbf{\bibinfo{volume}{403}},
  \bibinfo{pages}{1829--1833} (\bibinfo{year}{2010}).
\newblock \eprint{0912.3693}.

\bibitem{Bland-Hawthorn2016}
\bibinfo{author}{{Bland-Hawthorn}, J.} \& \bibinfo{author}{{Gerhard}, O.}
\newblock \bibinfo{title}{{The Galaxy in Context: Structural, Kinematic, and
  Integrated Properties}}.
\newblock \emph{\bibinfo{journal}{Annual Review of Astronomy \& Astrophysics}}
  \textbf{\bibinfo{volume}{54}}, \bibinfo{pages}{529--596}
  (\bibinfo{year}{2016}).
\newblock \eprint{1602.07702}.

\bibitem{Li2018ApJ...866...22L}
\bibinfo{author}{{Li}, T.~S.} \emph{et~al.}
\newblock \bibinfo{title}{{The First Tidally Disrupted Ultra-faint Dwarf
  Galaxy?: A Spectroscopic Analysis of the Tucana III Stream}}.
\newblock \emph{\bibinfo{journal}{\apj}} \textbf{\bibinfo{volume}{866}},
  \bibinfo{pages}{22} (\bibinfo{year}{2018}).
\newblock \eprint{1804.07761}.

\bibitem{Feroz2008MNRAS.384..449F}
\bibinfo{author}{{Feroz}, F.} \& \bibinfo{author}{{Hobson}, M.~P.}
\newblock \bibinfo{title}{{Multimodal nested sampling: an efficient and robust
  alternative to Markov Chain Monte Carlo methods for astronomical data
  analyses}}.
\newblock \emph{\bibinfo{journal}{\mnras}} \textbf{\bibinfo{volume}{384}},
  \bibinfo{pages}{449--463} (\bibinfo{year}{2008}).
\newblock \eprint{0704.3704}.

\bibitem{Feroz2009MNRAS.398.1601F}
\bibinfo{author}{{Feroz}, F.}, \bibinfo{author}{{Hobson}, M.~P.} \&
  \bibinfo{author}{{Bridges}, M.}
\newblock \bibinfo{title}{{MULTINEST: an efficient and robust Bayesian
  inference tool for cosmology and particle physics}}.
\newblock \emph{\bibinfo{journal}{\mnras}} \textbf{\bibinfo{volume}{398}},
  \bibinfo{pages}{1601--1614} (\bibinfo{year}{2009}).
\newblock \eprint{0809.3437}.

\bibitem{2017ApJ...835...77M}
\bibinfo{author}{{Marigo}, P.} \emph{et~al.}
\newblock \bibinfo{title}{{A New Generation of PARSEC-COLIBRI Stellar
  Isochrones Including the TP-AGB Phase}}.
\newblock \emph{\bibinfo{journal}{\apj}} \textbf{\bibinfo{volume}{835}},
  \bibinfo{pages}{77} (\bibinfo{year}{2017}).
\newblock \eprint{1701.08510}.

\bibitem{orphan_lmc_modelling}
\bibinfo{author}{{Erkal}, D.} \emph{et~al.}
\newblock \bibinfo{title}{{The total mass of the Large Magellanic Cloud from
  its perturbation on the Orphan stream}}.
\newblock \emph{\bibinfo{journal}{\mnras}} \textbf{\bibinfo{volume}{487}},
  \bibinfo{pages}{2685--2700} (\bibinfo{year}{2019}).
\newblock \eprint{1812.08192}.

\bibitem{gibbons_etal_2014}
\bibinfo{author}{{Gibbons}, S.~L.~J.}, \bibinfo{author}{{Belokurov}, V.} \&
  \bibinfo{author}{{Evans}, N.~W.}
\newblock \bibinfo{title}{{`Skinny Milky Way please', says Sagittarius}}.
\newblock \emph{\bibinfo{journal}{\mnras}} \textbf{\bibinfo{volume}{445}},
  \bibinfo{pages}{3788--3802} (\bibinfo{year}{2014}).
\newblock \eprint{1406.2243}.

\bibitem{mcmillan_2017}
\bibinfo{author}{{McMillan}, P.~J.}
\newblock \bibinfo{title}{{The mass distribution and gravitational potential of
  the Milky Way}}.
\newblock \emph{\bibinfo{journal}{\mnras}} \textbf{\bibinfo{volume}{465}},
  \bibinfo{pages}{76--94} (\bibinfo{year}{2017}).
\newblock \eprint{1608.00971}.

\bibitem{dehnen_binney_1998}
\bibinfo{author}{{Dehnen}, W.} \& \bibinfo{author}{{Binney}, J.}
\newblock \bibinfo{title}{{Mass models of the Milky Way}}.
\newblock \emph{\bibinfo{journal}{\mnras}} \textbf{\bibinfo{volume}{294}},
  \bibinfo{pages}{429--438} (\bibinfo{year}{1998}).
\newblock \eprint{astro-ph/9612059}.

\bibitem{hernquist_1990}
\bibinfo{author}{{Hernquist}, L.}
\newblock \bibinfo{title}{{An Analytical Model for Spherical Galaxies and
  Bulges}}.
\newblock \emph{\bibinfo{journal}{\apj}} \textbf{\bibinfo{volume}{356}},
  \bibinfo{pages}{359} (\bibinfo{year}{1990}).

\bibitem{emcee}
\bibinfo{author}{{Foreman-Mackey}, D.}, \bibinfo{author}{{Hogg}, D.~W.},
  \bibinfo{author}{{Lang}, D.} \& \bibinfo{author}{{Goodman}, J.}
\newblock \bibinfo{title}{{emcee: The MCMC Hammer}}.
\newblock \emph{\bibinfo{journal}{\pasp}} \textbf{\bibinfo{volume}{125}},
  \bibinfo{pages}{306} (\bibinfo{year}{2013}).
\newblock \eprint{1202.3665}.

\bibitem{amorisco_gmcs}
\bibinfo{author}{{Amorisco}, N.~C.}, \bibinfo{author}{{G{\'o}mez}, F.~A.},
  \bibinfo{author}{{Vegetti}, S.} \& \bibinfo{author}{{White}, S. D.~M.}
\newblock \bibinfo{title}{{Gaps in globular cluster streams: giant molecular
  clouds can cause them too}}.
\newblock \emph{\bibinfo{journal}{\mnras}} \textbf{\bibinfo{volume}{463}},
  \bibinfo{pages}{L17--L21} (\bibinfo{year}{2016}).
\newblock \eprint{1606.02715}.

\bibitem{erkal_etal_2017}
\bibinfo{author}{{Erkal}, D.}, \bibinfo{author}{{Koposov}, S.~E.} \&
  \bibinfo{author}{{Belokurov}, V.}
\newblock \bibinfo{title}{{A sharper view of Pal 5's tails: discovery of stream
  perturbations with a novel non-parametric technique}}.
\newblock \emph{\bibinfo{journal}{\mnras}} \textbf{\bibinfo{volume}{470}},
  \bibinfo{pages}{60--84} (\bibinfo{year}{2017}).
\newblock \eprint{1609.01282}.

\bibitem{pearson_pal5}
\bibinfo{author}{{Pearson}, S.}, \bibinfo{author}{{Price-Whelan}, A.~M.} \&
  \bibinfo{author}{{Johnston}, K.~V.}
\newblock \bibinfo{title}{{Gaps and length asymmetry in the stellar stream
  Palomar 5 as effects of Galactic bar rotation}}.
\newblock \emph{\bibinfo{journal}{Nature Astronomy}}
  \textbf{\bibinfo{volume}{1}}, \bibinfo{pages}{633--639}
  (\bibinfo{year}{2017}).
\newblock \eprint{1703.04627}.

\bibitem{banik_bovy_2019}
\bibinfo{author}{{Banik}, N.} \& \bibinfo{author}{{Bovy}, J.}
\newblock \bibinfo{title}{{Effects of baryonic and dark matter substructure on
  the Pal 5 stream}}.
\newblock \emph{\bibinfo{journal}{\mnras}} \textbf{\bibinfo{volume}{484}},
  \bibinfo{pages}{2009--2020} (\bibinfo{year}{2019}).
\newblock \eprint{1809.09640}.

\bibitem{agama}
\bibinfo{author}{{Vasiliev}, E.}
\newblock \bibinfo{title}{{AGAMA: action-based galaxy modelling architecture}}.
\newblock \emph{\bibinfo{journal}{\mnras}} \textbf{\bibinfo{volume}{482}},
  \bibinfo{pages}{1525--1544} (\bibinfo{year}{2019}).
\newblock \eprint{1802.08239}.

\bibitem{2019AJ....158..223P}
\bibinfo{author}{{Price-Whelan}, A.~M.} \emph{et~al.}
\newblock \bibinfo{title}{{Kinematics of the Palomar 5 Stellar Stream from RR
  Lyrae Stars}}.
\newblock \emph{\bibinfo{journal}{\aj}} \textbf{\bibinfo{volume}{158}},
  \bibinfo{pages}{223} (\bibinfo{year}{2019}).

\bibitem{2019MNRAS.485.4726K}
\bibinfo{author}{{Koposov}, S.~E.} \emph{et~al.}
\newblock \bibinfo{title}{{Piercing the Milky Way: an all-sky view of the
  Orphan Stream}}.
\newblock \emph{\bibinfo{journal}{\mnras}} \textbf{\bibinfo{volume}{485}},
  \bibinfo{pages}{4726--4742} (\bibinfo{year}{2019}).
\newblock \eprint{1812.08172}.

\bibitem{corner}
\bibinfo{author}{Foreman-Mackey, D.}
\newblock \bibinfo{title}{corner.py: Scatterplot matrices in python}.
\newblock \emph{\bibinfo{journal}{The Journal of Open Source Software}}
  \textbf{\bibinfo{volume}{24}} (\bibinfo{year}{2016}).

\end{thebibliography}
\bibliographystyleme{naturemag}
\bibliographyme{refs}

\begin{addendum}
 \item  
 Based in part on data acquired through the Australian Astronomical Observatory, under program A/2018B/09. We acknowledge the traditional owners of the land on which the AAT stands, the Gamilaraay people, and pay our respects to elders past, present and emerging.
 
 We thank Paul McMillan for providing the posterior chains for his fit to the Milky Way potential\citeme{mcmillan_2017}.
 
This project used public archival data from the Dark Energy Survey (DES). Funding for the DES Projects has been provided by the DOE and NSF (USA), MISE (Spain), STFC (UK), HEFCE (UK), NCSA (UIUC), KICP (U. Chicago), CCAPP (Ohio State), MIFPA (Texas A\&M), CNPQ, FAPERJ, FINEP (Brazil), MINECO (Spain), DFG (Germany) and the collaborating institutions in the Dark Energy Survey, which are Argonne Lab, UC Santa Cruz, University of Cambridge, CIEMAT-Madrid, University of Chicago, University College London, DES-Brazil Consortium, University of Edinburgh, ETH Z{\"u}rich, Fermilab, University of Illinois, ICE (IEEC-CSIC), IFAE Barcelona, Lawrence Berkeley Lab, LMU M{\"u}nchen and the associated Excellence Cluster Universe, University of Michigan, NOAO, University of Nottingham, Ohio State University, OzDES Membership Consortium, University of Pennsylvania, University of Portsmouth, SLAC National Lab, Stanford University, University of Sussex, and Texas A\&M University.

Based in part on observations at Cerro Tololo Inter-American Observatory, National Optical Astronomy Observatory, which is operated by the Association of Universities for Research in Astronomy (AURA) under a cooperative agreement with the National Science Foundation.

This work has made use of data from the European Space Agency (ESA) mission
{\it Gaia} (\url{https://www.cosmos.esa.int/gaia}), processed by the {\it Gaia}
Data Processing and Analysis Consortium (DPAC,
\url{https://www.cosmos.esa.int/web/gaia/dpac/consortium}). Funding for the DPAC
has been provided by national institutions, in particular the institutions
participating in the {\it Gaia} Multilateral Agreement.

Parts of this research were conducted by the Australian Research Council Centre of Excellence for All Sky Astrophysics in 3 Dimensions (ASTRO 3D), through project number CE170100013. 
 
 ZW is supported by a Dean’s International Postgraduate Research Scholarship at the University of Sydney.
 DM is supported by an Australian Research Council (ARC) Future Fellowship (FT160100206). JDS, SLM and DZ acknowledge the support of the Australian Research Council through Discovery Project grant DP180101791. 
 TSL and APJ are supported by NASA through Hubble Fellowship grant HST-HF2-51439.001 and HST-HF2-51393.001 respectively, awarded by the Space Telescope Science Institute, which is operated by the Association of Universities for Research in Astronomy, Inc., for NASA, under contract NAS5-26555.
 \item[Author Contributions] The \SSSSS\ program was initiated by TSL, DZ, KK and GFL, while 
 survey design and target selection for \SSSSS\ was undertaken by TSL and NS, and observations with the AAT were performed by GFL, KK, DM, SLM, JDS, DBZ, GDC and ZW.
 Data reduction, calibration and analysis was undertaken by SEK, TSL, APJ, ZW and GFL, and DE performed the dynamical analysis including stream fitting, orbit determination, and action comparison.
 All authors were involved in the discussion and interpretation of the results presented, and all contributed to the writing of the paper.
 \item[Competing Interests] The authors declare that they have no competing financial interests.
 \item[Correspondence] Correspondence and requests for materials
should be addressed to GFL \\
(email: geraint.lewis@sydney.edu.au).
\end{addendum}

\end{document}